\documentclass[aps,hyperref,twocolumn,floatfix,showpacs,superscriptaddress]{revtex4} 

\usepackage{graphicx}

\newcommand{\Hsp}{\hat{H}_{sp}}

\newcommand{\GPE}{Gross-Pitaevskii equation }

\newcommand{\Cnl}{C_{\rm nl}}

\begin{document}

\title{Simulations of thermal Bose fields in the classical limit}
\author{M.~J.~Davis}
\email{mdavis@physics.uq.edu.au}
\affiliation{Department of Physics, University of Queensland, St Lucia, QLD 
4072, Australia}
\affiliation{Clarendon Laboratory, Department of Physics, University of Oxford,
Oxford OX1~3PU, United Kingdom}
\author{S.~A.~Morgan}
\affiliation{Clarendon Laboratory, Department of Physics, University of Oxford,
Oxford OX1~3PU, United Kingdom}
\affiliation{Department of Physics and Astronomy, University College London, 
Gower Street, London WC1E~6BT, United Kingdom}
\author{K.~Burnett}
\affiliation{Clarendon Laboratory, Department of Physics, University of Oxford,
Oxford OX1~3PU, United Kingdom}

\begin{abstract}

We demonstrate that the time-dependent projected Gross-Pitaevskii
equation (GPE) derived earlier [Davis {\em et al.}, J. Phys. B
\textbf{34}, 4487 (2001)] can represent the highly occupied modes of a
homogeneous, partially-condensed Bose gas.  Contrary to the often held
belief that the GPE is valid only at zero temperature, we find that this
equation will evolve randomised initial wave functions to a state
describing thermal equilibrium.

In the case of small interaction strengths or low temperatures, our
numerical results can be compared to the predictions of Bogoliubov theory
and its perturbative extensions. This demonstrates the validity of the
GPE in these limits and allows us to assign a temperature to the
simulations unambiguously. 

However, the GPE method is non-perturbative, and we believe it can be used
to describe the thermal properties of a Bose gas even when Bogoliubov
theory fails. We suggest a different technique to measure the
temperature of our simulations in these circumstances. Using this approach we
determine the dependence of the condensate fraction and specific heat on
temperature for several interaction strengths, and observe the appearance of
vortex networks. Interesting behaviour near the critical point is observed and
discussed.

\end{abstract}

\maketitle

\section{Introduction}

The observation of Bose-Einstein condensation (BEC) in dilute alkali gases
\cite{JILA,MIT,RICE} heralds a new era in the study of quantum fields.  It
offers a unique opportunity to carry out experiments in the laboratory for
which  theoretical calculations beginning from a microscopic
model of the system are tractable. However, such calculations are fraught with difficulties
at finite temperatures.  While equilibrium perturbation theories have had much
success \cite{sam,fedichev,giorgini} dynamical calculations often require
severe approximations to be made.

In Ref.~\cite{formalism} we developed an approximate formalism to describe the
dynamics of a thermal Bose condensate  based on the Gross-Pitaevskii equation
(GPE).  This description is valid when the low-lying modes of the system are
classical, satisfying the criterion $N_k \gg 1$.  This is analogous to the
situation in laser physics, where the highly occupied laser modes can be well
described by classical equations.  We proceeded by dividing the field operator
into a classical region represented by  a wave function $\psi(\mathbf{x})$
describing the condensate and its coherent excitations,  with the
remainder of the field described by the quantum operator
$\hat{\eta}({\bf x})$.  We derived an equation of motion for $\psi(\mathbf{x})$
that we called the finite temperature Gross-Pitaevskii
equation (FTGPE).

 The FTGPE is a rather complicated equation, however, and in Ref.~\cite{ftgpe}
we briefly described the first results from the simpler projected
Gross-Pitaevkskii equation (PGPE) obtained by neglecting the operator
$\hat{\eta}({\bf x})$. These results demonstrate that the GPE alone can
represent thermal Bose gases.  In this paper we elaborate on these results and
describe our method in more detail. We also consider the effect of strong
particle interactions on the thermal distributions and investigate the
appearance of vortices in our simulations. 

The use of the dynamical GPE at finite temperature was originally proposed by
Svistunov, Kagan, and co-workers \cite{boris,kagan1,kagan2,kagan3}.  Despite
this suggestion first appearing in 1991, there have been relatively few
numerical studies based on this approach.  Damle {\em et al.}\ have performed
calculations of the approach to equilibrium of a near ideal superfluid
\cite{damle}, while Marshall {\em et al.} \cite{Marshall} carried out a
qualitative study of evaporative cooling using a 2D GPE.  References 
\cite{goral1,goral2,sinatra1,sinatra2,sinatra3,bijlsma} also use classical methods to
represent thermal Bose-condensed systems. Similar approximations to other
quantum field equations have been successful elsewhere \cite{Turok}.

This paper is organised as follows.  In Sec.~\ref{outline} we give a brief
derivation of the finite temperature Gross-Pitaevskii equation.  In 
Sec.~\ref{pgpe} we describe and justify the simplification of the FTGPE to
the projected Gross-Pitaevskii equation, before describing the simulations we
have carried out in Sec.~\ref{simulations}.  Section~\ref{sec:equil_evidence} presents
the qualitative evidence that the simulations have reached equilibrium, while
Sec.~\ref{quantitative} carries out a quantitative analysis of our numerical
data.   Section~\ref{frac_vortex} discusses the behaviour of the condensate
fraction, specific heat, and vorticity of the system with temperature, before we conclude in 
Section~\ref{conclusions}.

\section{Outline of formalism}\label{outline}

A full derivation of the FTGPE and a discussion of the physics described by each
of the terms can be found in Ref.~\cite{formalism}.  Here we outline the
derivation beginning with the equation of motion
for the Bose field operator 
\begin{equation}
i \hbar\frac{\partial \hat{\Psi}({\bf x})}{\partial t} =
\Hsp \hat{\Psi}({\bf x}) + 
U_0
\hat{\Psi}^{\dag}({\bf x})
\hat{\Psi}({\bf x})\hat{\Psi}({\bf x}),
\label{eqn:field_operator}
\end{equation}
where $U_0 = 4\pi \hbar^2 a / m$ is the effective interaction
strength at low momenta,  $a$ is the {\em s}-wave scattering length, and $m$ is
the particle mass.  $\Hsp$ is the single-particle hamiltonian defined by
\begin{equation}
\Hsp = -\frac{\hbar^2}{2m} \nabla^2 + V_{\mathrm{trap}}({\bf x}),
\end{equation}
where $V_{\mathrm{trap}}({\bf x})$ is the external trapping potential, if any is present.

 The route to the usual GPE is to assume that the full field
operator can be replaced by a wave function $\psi(\mathbf{x})$---i.e.\ that {\em
all} quantum fluctuations can be neglected.  We proceed instead by defining a
projection operator $\hat{\mathcal{P}}$ such that
\begin{eqnarray}
\hat{\mathcal{P}}\hat{\Psi}({\bf x})  = 
\sum_{{\bf k} \in C} \hat{a}_{\bf k} \phi_{\bf k}({\bf x}),
\end{eqnarray}
where the region $C$ is {\em determined} by the 
requirement that $\langle \hat{a}_{\bf k}^{\dag} \hat{a}_{\bf k}\rangle \gg 1$,
and the set $\{\phi_{\bf k}\}$ defines some basis in which the field operator is
approximately diagonal at the boundary of $C$.
For these modes, the quantum fluctuation part of the projected field operator can be
ignored, and so we replace $\hat{a}_{\bf k} \rightarrow c_{\bf k}$ and write
\begin{eqnarray}
\psi(\mathbf{x})  = \sum_{{\bf k} \in C} c_{\bf k} \phi_{\bf k}({\bf x}).
\end{eqnarray}

Defining the operator $\hat{\mathcal{Q}} = \hat{\openone} - \hat{\mathcal{P}}$ and 
$\hat{\mathcal{Q}}\hat{\Psi}({\bf x}) = \hat{\eta}({\bf x})$, operating on
Eq.~(\ref{eqn:field_operator}) with $\hat{\mathcal{P}}$ and taking the mean
value results in what we call the finite temperature GPE
\begin{eqnarray}
i \hbar \frac{\partial {\psi}({\bf x})}{\partial t}
&=& 
\Hsp \psi(\mathbf{x}) + U_0
\hat{\mathcal{P}}\left\{|\psi(\mathbf{x})|^2 \psi(\mathbf{x})\right\} 
\nonumber
\\
&+&
U_0\hat{\mathcal{P}}\left\{2|\psi(\mathbf{x})|^2
\langle \hat{\eta}({\bf x})\rangle
+
\psi(\mathbf{x})^2 \langle\hat{\eta}^{\dag}({\bf x})\rangle \right\}
\nonumber
\\
&+&
U_0\hat{\mathcal{P}}\left\{\psi^*({\bf x})\langle
\hat{\eta}({\bf x})\hat{\eta}({\bf x})\rangle 
+
2\psi(\mathbf{x})\langle\hat{\eta}^{\dag}({\bf x})\hat{\eta}({\bf x})\rangle \right\}
\nonumber
\\
&+&
U_0\hat{\mathcal{P}}\left\{\langle\hat{\eta}^{\dag}({\bf x})
\hat{\eta}({\bf x})\hat{\eta}({\bf x})\rangle 
\right\},
\label{eqn:ftgpe}
\end{eqnarray}
This equation describes the full dynamics of the coherent region and its
coupling to an effective heat bath described by $\hat{\eta}({\bf x})$. In
general, the non-equilibrium evolution depends on the coupling between these
two regions and the exchange of energy and particles that this allows. The
FTGPE must be complemented by an equation of motion for $\hat{\eta}({\bf x})$
and in principle this can be obtained using a form of quantum kinetic theory.

The only approximation that has been made in the derivation of the FTGPE is
that the modes represented by $\psi({\bf x})$ must satisfy the criterion of
classicality, that is $N_k \gg 1$.  The FTGPE is a non-perturbative equation,
and therefore we expect that it will be valid in the region of the phase
transition as long as only the highly occupied modes are treated.  There is 
perhaps a misperception in the BEC community that the GPE is only valid at
$T=0$.   However, it is well known that close to the phase transition a
classical description of the longer length scales involved is completely
appropriate. This is exactly what the GPE describes, and in fact it has been
used as a model of phase transitions in other areas of condensed matter
physics.  Indeed our model has the same energy functional for these modes as
used in the classical renormalization group theory of the superfluid phase
transition.  It therefore seems reasonable to expect that the same
approximations are valid in this case.

The  physical processes described by the various terms of
Eq.~(\ref{eqn:ftgpe})  are discussed in detail in  Ref.~\cite{formalism}.  In
this paper we concentrate on a simplification of the FTGPE which is effectively
a model of a restricted system.  This allows us to demonstrate some of the
properties of the GPE without having to solve the more complicated equation.

\section{The projected GPE}\label{pgpe}

In this paper,  we wish to show that the GPE {\em alone} can describe evolution
of general configurations of the coherent region  $C$  towards an equilibrium
that can be parameterised by a temperature. We therefore ignore all terms
involving  $\hat{\eta}({\bf x})$ in Eq.~(\ref{eqn:ftgpe}) and concentrate on
the first line 
\begin{equation}
i \hbar \frac{\partial {\psi}({\bf x})}{\partial t}
=
\Hsp \psi(\mathbf{x}) + U_0
\hat{\mathcal{P}}\left\{|\psi(\mathbf{x})|^2 \psi(\mathbf{x})\right\},
\label{eqn:pgpe}
\end{equation}
which we call the projected Gross-Pitaevskii equation (PGPE).
  Although Eq.~(\ref{eqn:pgpe}) is
completely reversible, it is well known that deterministic nonlinear systems
with only a few degrees of freedom exhibit chaotic, and hence ergodic behaviour
\cite{reichl}.  If many modes are occupied, the PGPE contains many degrees 
of freedom and it is therefore  reasonable to expect it to evolve 
to equilibrium (except for specially chosen initial conditions such as
 eigenstate solutions).

The projected GPE describes a microcanonical system.  However, if the region
$C$ is large, then its fluctuations in energy and particle number in the  grand
canonical ensemble would be small.  Hence we expect the final  equilibrium
state of the projected GPE to be similar to that of the finite temperature GPE
coupled to a bath $\hat{\eta}({\bf x})$ with the appropriate chemical potential
and temperature.  The detailed non-equilibrium dynamics of the system {\em
will} depend on the exchange of energy and particles between $C$ and the
bath---however, we leave the coupling of  $\psi(\mathbf{x})$ and
$\hat{\eta}({\bf x})$ to be addressed in future work.

\subsection{The projector}
The spatial representation of the projection operation is written
\begin{eqnarray}
\hat{\mathcal{P}}\{F(\mathbf{x}) \}
%& =& \sum_{k \in C} |k\rangle\langle k|,\nonumber\\
&=& \sum_{k \in C}\phi_k({\bf x}) \int d^3{\bf x}' \;\phi_k^*({\bf x}')
F({\bf x}'),
\label{eqn:projector}
\end{eqnarray}
and this operation must be carried out numerically every time  we calculate the
nonlinear term in the PGPE.  This is  a very time consuming
operation in general, taking many times longer than calculating
$|\psi({\bf x})|^2 \psi({\bf x})$ itself.

The operation is much simpler numerically if we use a plane-wave basis in our
projector
\begin{equation}
 \phi_k({\bf x}) = 
\frac{\exp(i {\bf k} \cdot {\bf x})}{\sqrt{V}}
\end{equation}
where $V$ is the volume of our system.  In this case Eq.~(\ref{eqn:projector})
becomes simply the application of a forward Fourier transformation to our
function $F(\mathbf{x})$, followed by
an inverse Fourier transformation that includes only the modes in the
coherent region.  
Thus our numerical procedure is 
\begin{eqnarray}
\hat{\mathcal{P}}\{ F({\bf x})\}& =&
\mbox{IFFT}\bigg\{ P({\bf k}) \times
 \mbox{FFT} \left[F({\bf x})\right] \bigg\},
\end{eqnarray}
where FFT and IFFT refer to the forward and inverse fast Fourier transform 
operations respectively, and $P({\bf k})$ is the representation of the
projector $\hat{\mathcal{P}}$ in Fourier space.   There are very efficient
routines available to carry out FFTs, and so we find
that it is extremely advantageous numerically to define our projector in the
plane-wave basis.

\subsection{Implications} 

For any non-periodic trapping potential, the use of a plane-wave basis is at
odds with our requirement that the basis  must approximately diagonalise
$\psi(\mathbf{x})$ at the boundary of the region $C$.    In fact, it may not even
satisfy this requirement for a periodic potential if the boundary of the
coherent region occurs at a low enough energy.

If we consider a homogeneous system, however, the plane-wave basis will always
satisfy our requirements.  In this case  the effect of a condensate on the
excitations of the system is simply to mix modes of momenta ${\bf p}$ and
$-{\bf p}$.  Thus even if $\psi(\mathbf{x})$ is not diagonalised at the boundary
of $C$, we can still apply the projector cleanly in Fourier space.  For these
reasons, the simulations that we present in this paper are for the homogeneous
Bose gas. We intend to address the issue of projectors for the trapped Bose gas
in future work.

A direct advantage of simulating the homogeneous system is that the condensate
occupation is readily identified as the $\mathbf{k} = 0$ component of the wave
function.  This is in contrast to the trapped case, where the condensate mode
changes with the condensate fraction.  In general the condensate fraction must
be determined by diagonalisation, which can be a very time consuming procedure
\cite{goral2}.

\section{Simulations}\label{simulations}

We have performed simulations for a fully three-dimensional homogeneous Bose gas
with periodic boundary conditions.  
The dimensionless equation we compute is
\begin{equation}
i \frac{\partial \psi(\tilde{ \bf x})}{\partial \tau}
= -\tilde{\nabla}^2 \psi( \tilde{\bf x}) + 
C_{\rm nl} \hat{\mathcal{P}} \{|\psi(\tilde{ \bf x})|^2\psi(\tilde{ \bf x})\}, 
\label{eqn:gpe}
\end{equation}
where the normalisation of the wave function has been defined to be 
\begin{equation}
\int d^3\tilde{\bf x} \,|\psi(\tilde{\bf x})|^2 = 1.
\end{equation}  
The nonlinear constant is 
\begin{equation}
C_{\rm nl} = \frac{2 m N U_0}{\hbar^2 L},
\end{equation} 
where $N$ is the total number of particles in the system, and $L$ is the
period. Our dimensionless parameters are $\tilde{\bf x} = {\bf x}/L$, wave
vector  $\tilde{\bf k} = {\bf k} L$, energy  $\tilde{\varepsilon} = \varepsilon
/ \varepsilon_L$, and  time $\tau = \varepsilon_L t/ \hbar$, with 
$\varepsilon_{L} = \hbar^2/(2 m L^2)$.

\subsection{Parameters}
The two parameters that determine all properties of the system are the
projector $\hat{\mathcal{P}}$ and the nonlinear constant $\Cnl$. 

\subsubsection{Projector $\hat{\mathcal{P}}$}

We have chosen a projection operator such that  all modes included in the
simulations have $|{\bf k}| < 15 \times 2\pi/L$, which enables us to use a
computationally efficient numerical grid of $32 \times 32 \times 32$ points.
This means that 13997 modes are included in the system.

\paragraph*{Grid size and aliasing}

The nonlinear term of the GPE can generate momentum components up to three
times larger than those which exist in the original wave function.  Thus it
would seem that calculating the term $|\psi(\mathbf{x})|^2 \psi(\mathbf{x})$ on a
grid only slightly larger than the projector would cause problems with
aliasing.  The correct procedure would be instead to calculate this term on a
grid size of  $96 \times 96 \times 96$ points before performing the projection
operation.

To check the effect of grid size we have performed simulations where the
nonlinear term was calculated on grids of size 32, 64, and 96 points, and found
that there is no difference in the equilibrium properties of the system.  The
detailed dependence of the condensate population during evolution is 
different in detail for each size grid, but follows the same average curve.  The
same behaviour is observed when adjusting the accuracy parameter of our
adaptive step size algorithm  for evolving the GPE.

We attribute this behaviour to the deterministic chaos exhibited by the
system.  Any small numerical error is eventually magnified such that the system
follows a quite different microscopic path through phase space, although 
the resulting macroscopic (average) properties are unaffected.

\subsubsection{Nonlinearity $C_{\rm nl}$}

We note that the choice of the nonlinear constant  determines only  the
ratio of $N U_0 / L$. This means that for a given value of $C_{\rm nl}$, we can
choose the  parameters $N$, $U_0$ and $L$ such that our condition $N_{\bf k}
\equiv N|c_{\bf k}|^2 \gg 1$ is always satisfied for a given physical
situation.

We have performed three series of simulations with
nonlinearities of $C_{\rm nl} = 500$, $2000$, and $10000$.  The highest
value of $C_{\rm nl}$ was chosen such that all
the states contained in the calculation are phonon-like
for a large condensate fraction.  The boundary between phonon-like and
particle-like states for the homogeneous gas is 
\begin{equation}
\frac{\hbar^2 k_0^2}{2 m} = n_0 U_0,
\label{eqn:k0}
\end{equation}
where  we have defined $N_0$ to be the condensate number within
the volume $L^3$, and thus $n_0 = N_0 / L^3$ is the condensate density.
Converting Eq.~(\ref{eqn:k0}) to dimensionless units we find that 
\begin{equation}
\tilde{k}_0 = \sqrt{C_{\rm nl}\frac{N_0}{N}},
\label{eqn:k0_dim}
\end{equation}
and therefore for a condensate fraction of $N_0 / N = 1$ we have
\begin{eqnarray}
C_{\rm nl} = 10000 
& \rightarrow & \tilde{k}_0 \approx 15.9 \times 2 \pi,\nonumber\\
C_{\rm nl} = 2000 
& \rightarrow & \tilde{k}_0 \approx 7.12 \times 2 \pi,\nonumber\\
C_{\rm nl} = 500 
& \rightarrow & \tilde{k}_0 \approx 3.56 \times 2 \pi.\nonumber
\end{eqnarray}
We find that computations with smaller values of $C_{\rm nl}$ take
comparatively longer to reach equilibrium.  This is because the equilibration
rate is approximately proportional to $C_{\rm nl}^2$, whereas the minimum
time-step required for a given accuracy in the numerical integration of the
PGPE only increases slowly with decreasing  $\Cnl$.

To give an indication of how these dimensionless parameters compare to
experimental setups, for $\Cnl = 10000$ we can choose $^{87}$Rb atoms with
$N=1.8\times10^6$ and $L\approx26$ $\mu$m to give a number density of about
$10^{14}$ cm$^{-3}$---similar to current experiments on BEC in traps.

\subsection{Initial wave functions}

We begin our simulations with  strongly non-equilibrium wave functions with a
chosen total energy $\tilde{E}$.  We construct these by populating the amplitudes of the
wave function components $c_{\bf k}$ in the expansion
\begin{equation}
\psi({\bf x},0) = \sum_{{\bf k} \in C} c_{\bf k} e ^{i {\bf k}\cdot {\bf x}}.
\label{eqn:wave_expanded}
\end{equation}
The populations $|c_{\bf k}|^2$ are chosen such that the distribution is as flat as
possible, while the phases of the amplitudes are chosen at random \cite{kagan3}. 

The total energy $\tilde{E}$ is a constraint on 
 the distribution of amplitudes.
The energy of  a pure condensate is $\tilde{E}_0 = C_{\rm nl}/2$,
all of this being due to interactions---the kinetic energy is zero. 
To have a wave function with
an energy not much larger than  $C_{\rm nl}/2$, the occupations of the
$\tilde{k}=0$ state and the $\tilde{k}= 2 \pi$ states cannot be equal.  (We use
the notation $\tilde{k} \equiv |\tilde{{\bf k}}|$.)
Therefore, for the lowest energy simulations the initial condensate population
is necessarily larger than the excited state populations.

\begin{figure}\centering
\includegraphics{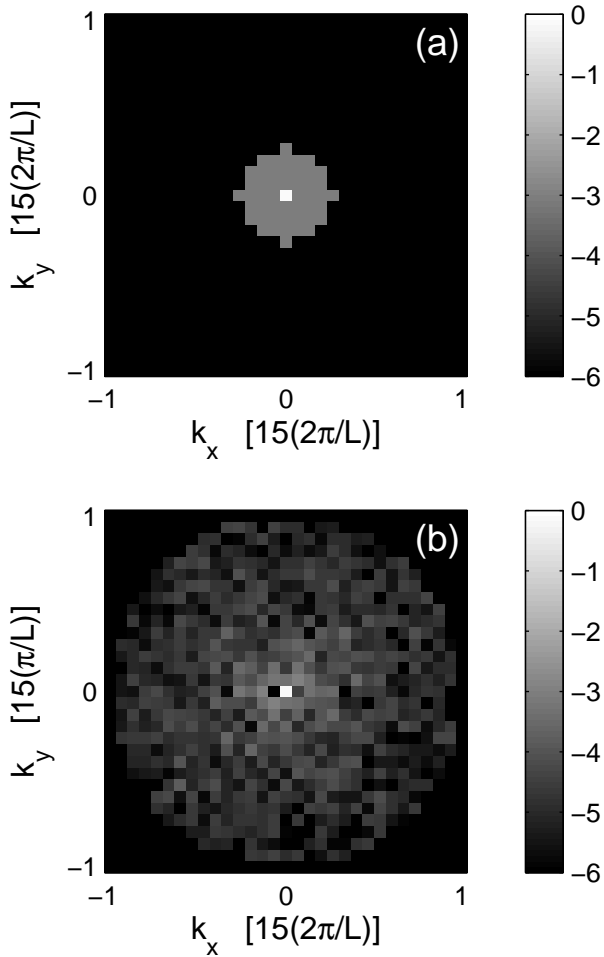}
\caption{
Two dimensional slices of wave functions through the $k_z = 0$ plane in momentum space 
for the  $\Cnl = 10000$, $\tilde{E} =
7000$ simulations.  (a) Base 10 logarithm of the {\em k}-space wave function
 at $\tau = 0$ (b) Base 10 logarithm of the {\em k}-space wave function at $\tau = 0.2$ once
 the system has reached equilibrium.}
\label{fig:wave1} 
\end{figure}

To ensure that the initial wave functions are sufficiently randomised,
we enforce the condition that all 123 states with $\tilde{k} \leq 3 \times 2
\pi$ must have some initial population, while all other components may be
unoccupied.  For low energies, when this distribution including the condensate 
cannot be totally flat, we keep the populations of the components with $1 \leq \tilde{k}/2\pi
\leq 3$ equal, and adjust the condensate population such that the wave function
has the energy we require.  An example of this situation is shown in 
Fig.~\ref{fig:wave1}(a) for the $\tilde{E} = 7000$ initial wave function in the
$\Cnl = 10000$ simulation series.

For simulations with a  sufficiently high total energy $\tilde{E}$ that the
inner  123 components may have equal population, we continue to 
add further shells
of higher $k$ to our wave function. The amplitudes of the inner components
are readjusted  to maintain the required normalisation. 
This causes the energy of the system  to increase monotonically with
each new shell until  we find two wave functions that bound the energy we are
looking for,  differing only in their outermost shell.  We then adjust the
population of the  outermost shell downwards until we reach the required
energy.  

This procedure is necessary due to the nonlinearity of the problem.  
In the case of the ideal gas
($C_{\rm nl} = 0$),
we can calculate the kinetic energy (and hence the total energy) of the wave
function simply by knowing the
distribution of $|c_{\bf k}|^2$, via
\begin{eqnarray}
{E}_{\rm kin} &=& -\frac{\hbar^2}{2m}\int d^3{\bf x}\, \psi^*({\bf x})
{\nabla}^2 \psi(\mathbf{x}),\nonumber\\
&=&\frac{\hbar^2}{2m}
\sum_{\bf k} |c_{\bf k}|^2 k^2.
\label{eqn:kin_energy}
\end{eqnarray}
However, for $C_{\rm nl} > 0$ we must also add the
interaction energy of the wave function to the total energy.  This is
\begin{eqnarray}
{E}_{\rm int} &=& \frac{U_0}{2}  \int d^3{\bf x} \,
|\psi(\mathbf{x})|^4,\nonumber\\
&=& \frac{U_0}{2} \sum_{\bf p q mn} 
c_{\bf p}^* c_{\bf q}^* c_{\bf m} c_{\bf n} 
\delta_{{\bf p} + {\bf q} - {\bf m} - {\bf n}},
\label{eqn:int_energy}
\end{eqnarray}
and depends nontrivially on the $\{c_{\bf k}\}$. 

\begin{figure}\centering
\includegraphics{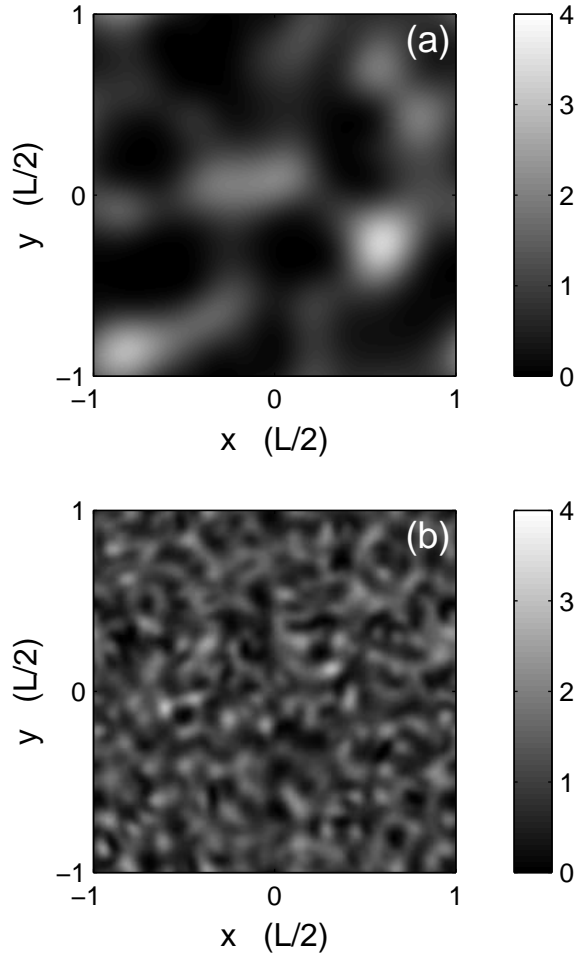}
\caption{
Two dimensional slices of wave functions near the $z = 0$ plane in real space 
for the  $\Cnl = 10000$, $\tilde{E} =
7000$ simulation.  (a) Base 10 logarithm of the  real-space wave function
 at $\tau = 0$ (b) Base 10 logarithm of the real-space wave function at $\tau = 0.2$ once
 the system has reached equilibrium.}
\label{fig:wave2}
\end{figure}

Further images of initial and final state wave functions are shown in
Fig.~\ref{fig:wave1} in \emph{k}-space, and Fig.~\ref{fig:wave2} in real space.

\subsection{Evolution}\label{sec:length}

The PGPE is evolved in the interaction picture,  using a fourth-order
Runge-Kutta method with adaptive step size determined by estimating the
fifth-order truncation error.  The
acceptable relative truncation error was set to be $10^{-10}$ for all
components with an occupation of $\ge 10^{-4} N_0/N$.  This resulted in typical
time steps as presented in Table~\ref{tab:time_steps}, which could be integrated
in a reasonable time on a modern workstation.

We evolve the initial wave functions for at least twice as long as it takes for
the system to reach equilibrium, based on the observation of the behaviour of
the condensate fraction (see Sec.~\ref{sec:equil_evidence}). The time period for each value
of $\Cnl$ is also given in  Table~\ref{tab:time_steps}. Thus the longest of
these simulations required $\sim 5 \times 10^5$ time steps.

\begin{table}\centering
\begin{tabular}{cccc}
$\qquad\Cnl\qquad$ &
\begin{tabular}{c}
Min. time step\\ ($10^{-6}$)
\end{tabular}&
\begin{tabular}{c}
Max. time step\\ ($10^{-6}$)
\end{tabular}&
\begin{tabular}{c}
Length of \\
evolution $\tau$
\end{tabular}
\\
\hline
500	&    4	&6 & 2.0\\
2000	&  1.6 & 4.4 & 0.4\\
10000	&  0.45    & 1.2 & 0.2  \\
\end{tabular}
\caption{The typical minimum and maximum time steps for the simulations.  The
minimum is for high energy simulations, and the maximum for low energy.}
\label{tab:time_steps}
\end{table}

\section{Evidence for equilibrium}\label{sec:equil_evidence}

Although the PGPE is completely reversible, the final state wave functions
displayed in Figs.~\ref{fig:wave1} and~\ref{fig:wave2} indicate that the
simulations have evolved the system to an apparent equilibrium 
state. The
{\em k}-space distributions have evolved from initially being flat to a form
that is peaked at the centre, and tails away towards the edges.  Also, there is
a smoothing out of both the
phase and density profiles of the real-space wave function. 
After a certain time of evolution $\tau_{\rm eq}$, the plots for the wave
functions appear to be isomorphic for $\tau > \tau_{\rm eq}$.

We would like to note that the equilibrium properties depend only on the
total energy and momentum of the initial wave function---they are independent
of the shape of the initial distribution in $k$-space.  We have performed simulations
with non-spherical initial wave functions, and found that they evolve to a
spherical equilibrium state.  Also, as the GPE conserves momentum, for the
condensate to form in the $k=0$ mode the initial distribution must have zero
total momentum.   We have performed simulations where the initial
distribution had a finite momentum, and observed the condensate to form in a
non-zero momentum state as expected. 

To determine the properties of the system at equilibrium, in theory we should
carry out many different simulations each with the same initial  populations
but with different choices of the initial phases, and then take the ensemble
average. However, this is an extremely large computational task.  Instead, we
assume the ergodic theorem applies, such that the time average over the
evolution of a single system at equilibrium is equivalent to the ensemble
average over many  different systems. We therefore perform a time-average over
the last 50 wave functions saved, all with $\tau > \tau_{\rm eq}$.

\subsubsection{Condensate occupation}
 
\begin{figure}\centering
\includegraphics[width=8.6cm]{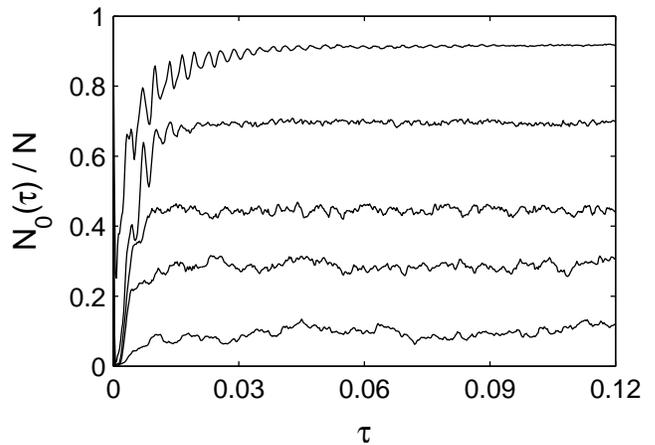}
\caption{Plot of the initial time evolution of $N_0(\tau) / N$ for four different simulation
energies with $\Cnl = 10000$. From top to bottom: $\tilde{E} = 5500, 7000,8500,
9250, 10000$.  The simulations were run until $\tau = 0.2$.
Other values of the nonlinearity give qualitatively similar results.}
\label{fig:n0_evolution}
\end{figure}

Strong evidence that the simulations have reached equilibrium is provided by
the time dependence of the condensate population.  For all simulations
this settles down to an average value (dependent on the energy $\tilde{E}$)
that fluctuates by a small amount.  
The initial  time evolution of the condensate fraction for five
different energies with  $\Cnl = 10000$ is shown in Fig.~\ref{fig:n0_evolution}. 

\begin{figure}\centering
\includegraphics[width=8.6cm]{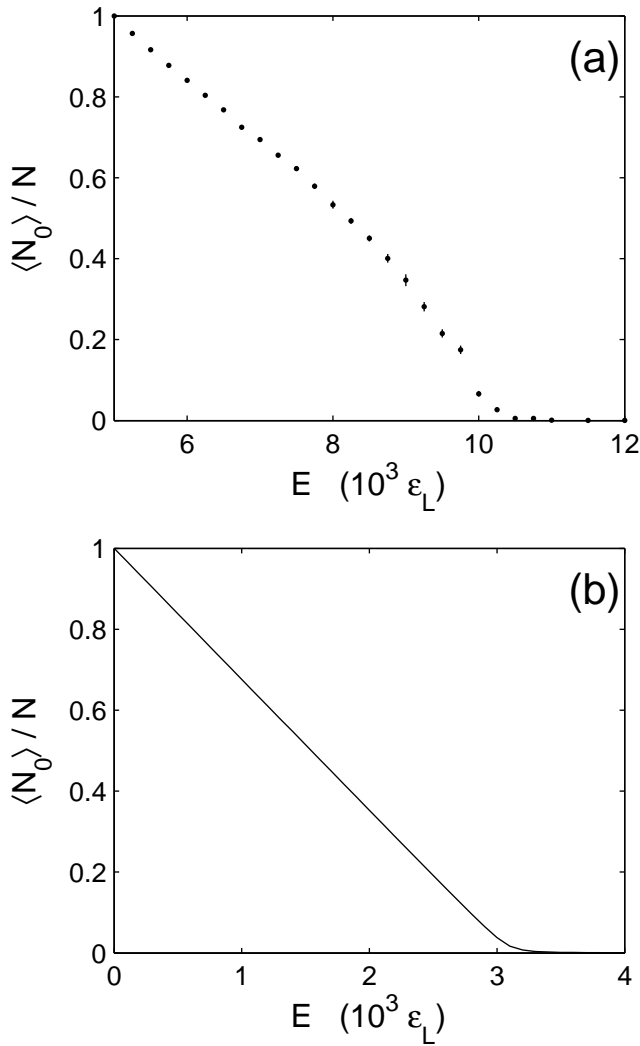}
\caption{(a) Condensate fraction plotted against total energy after each
individual simulation has reached equilibrium for $C_{\rm nl} = 10000$.  The
barely discernible
vertical lines on each point indicate the magnitude of the fluctuations.  (b) 
The curve for the same system, but calculated for the ideal gas.} 
\label{fig:n0ve}
\end{figure}

The average condensate occupation in equilibrium for all simulations for the
$\Cnl=10000$ case are presented in Fig.~\ref{fig:n0ve}(a). The fluctuations of
the condensate population are indicated by  the (barely visible) vertical
lines at each point, and these are largest for the $\tilde{E} = 9000$
simulation. For comparison, the corresponding curve for the ideal gas is
plotted in  Fig.~\ref{fig:n0ve}(b).  We can see that for $\Cnl = 0$ the curve
is linear up to the transition point, but the $\Cnl=10000$ curve displays a
distinct bulge in this region.  The shape of the corresponding curves for
$\Cnl=500$ and $2000$ fall in between the $\Cnl = 0$ and $10000$ cases.

\subsubsection{Particle distribution}

Further evidence of equilibrium is provided by the distribution of the
particles in momentum space.  Rather than using the plane-wave basis, we
transform the wave functions into the quasiparticle basis of quadratic
Bogoliubov theory.  For the homogeneous gas, this theory can be solved 
analytically and we can write the quasiparticle amplitude
$b_{\bf k}$ as 
\begin{equation}
b_{\bf k} =  u_{\bf k} c_{\bf k} - v_{\bf k} c_{-{\bf k}},
\label{homo_qp_transform}
\end{equation}
where
\begin{equation}
u_{\bf k} = \frac{1}{\sqrt{1 - \alpha_k^2}},
\qquad
v_{\bf k} = \frac{-\alpha_k}{\sqrt{1 - \alpha_k^2}},
\label{homo_uk_vk}
\end{equation}
and $\alpha_k$ is given by 
\begin{equation}
\alpha_k = 1 + y_k^2 - y_k\sqrt{2 + y_k^2}.
\label{eqn:alpha}
\end{equation}
In this last equation, the dimensionless wave vector $y_k$ is given by $y_k =
k/k_0$ with $k_0$ as defined in Eq.~(\ref{eqn:k0}). The normalisation
condition  $u_{\bf k}^2  - v_{\bf k}^2 = 1$ is automatically satisfied by
Eq.~(\ref{homo_uk_vk}). From Eq.~(\ref{eqn:k0_dim}) we can see that the sole
parameters of the transformation are the condensate fraction $\langle N_0
\rangle / N$, and the nonlinear constant  $\Cnl$.

\begin{figure}\centering
\includegraphics[width=8.6cm]{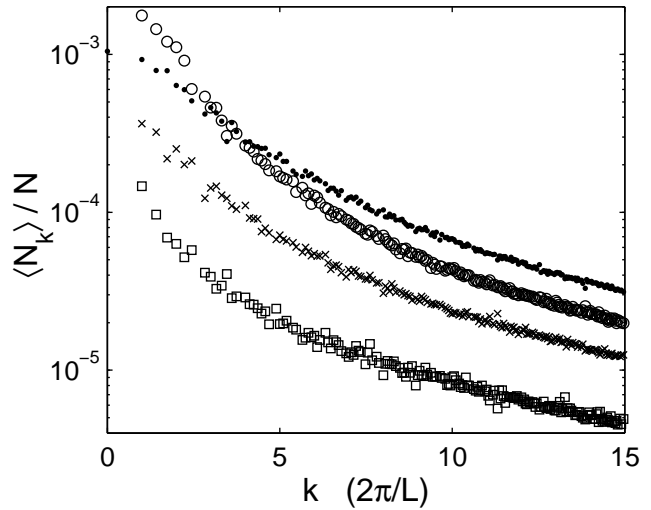}
\caption{Plots of the equilibrium Bogoliubov quasiparticle 
distributions averaged over time and angle for four different total energies. 
Squares $\tilde{E} = 6000$, crosses  $\tilde{E} = 7500$, circles 
$\tilde{E} = 9000$, dots $\tilde{E} = 11000$.  The mean condensate occupation 
for the first three distributions is off axis.} 
\label{fig:qp_distribution}
\end{figure}

We time average the populations of the quasiparticles states $N_{\bf k}/N = |b_{\bf
k}|^2$  as was described in Sec.~\ref{sec:equil_evidence} to give $\langle
N_{\bf k} \rangle/N$, and finally average over angle so that we can produce a one
dimensional plot of  $\langle N_{ k} \rangle/N$. 
This distribution for four different
simulation energies and $\Cnl = 10000$ is shown in
Fig.~\ref{fig:qp_distribution}. 

We can see that the shape of the curves is surprisingly smooth for each energy,
suggesting that the system is in equilibrium.  The plot of the distribution for
any individual wave function is scattered about the average.

We have also determined the fluctuations of the population of the
quasiparticle modes.  The grand canonical ensemble for the Bose gas
 predicts the relationship
\begin{equation}
\langle \Delta N_k \rangle^2  = \langle N_k\rangle^2 + \langle N_k\rangle,
\end{equation}
for $k \neq 0$, which  in the classical limit $\langle N_k \rangle
\gg 1$ gives 
\begin{equation}
\langle \Delta N_k \rangle \approx \langle N_k\rangle,
\end{equation}
This is  indeed the behaviour that we observe.  Although  we are evolving a
microcanonical system, in this case there are such a large number of modes that
the remainder of the system acts as a bath for any individual mode  and the
result still applies.

\section{Quantitative analysis of the distributions}\label{quantitative} 

While the data presented in Sec.~\ref{sec:equil_evidence} indicates that the
PGPE is evolving the system to equilibrium, as yet we have presented no
quantitative evidence.  To demonstrate conclusively that equilibrium has been
reached, we need to be able to assign a \emph{temperature} to the simulations. 
In this section we measure a temperature for a given simulation by comparing
the distribution function of the numerical simulations against a predicted
energy spectrum.

%One method of achieving this is to fit the simulation data to a predicted
%quasiparticle distribution, and this is how we proceed in the next section.

\subsection{Expected equilibrium distribution}

The GPE is the high occupation (classical) limit of the full equation for the
Bose field operator, Eq.~(\ref{eqn:field_operator}).  Therefore, in equilibrium
we expect the mean occupation of mode ${ k}$ to be the classical limit of the
Bose-Einstein  distribution---i.e.\ the equipartition relation

\begin{equation}
\langle N_{k} \rangle = \frac{k_B T}{\varepsilon_{ k} - \mu},
\label{eqn:occupation}
\end{equation}
where $k$ labels the \emph{eigenstates} of the system.  In general these will be
some type of quasiparticle mode.
Manipulating  Eq.~(\ref{eqn:occupation}), we find that
\begin{equation} 
{\varepsilon}_{k} = \frac{k_B T}{\langle N_k \rangle} + \mu.
\label{eqn:disperse_fit}
\end{equation}
The equilibrium condensate occupation according to the
equipartition relation will be given by Eq.~(\ref{eqn:occupation}) with 
$\langle N_k \rangle \rightarrow \langle N_0 \rangle$ and
$\varepsilon_k \rightarrow \lambda$ (the condensate eigenvalue).  From this expression we can solve for 
the chemical potential
\begin{equation}
\mu = \lambda - \frac{k_B T}{\langle N_0 \rangle}.
\end{equation}
Substituting this result into Eq.~(\ref{eqn:disperse_fit}), and converting
to dimensionless units we find
\begin{equation} 
\frac{\tilde{\varepsilon}_{k} - \tilde{\lambda}}{\tilde{T}} = \left(\frac{N}{\langle N_{ k} \rangle} -
\frac{N}{\langle N_0 \rangle}\right), 
\label{eqn:fit} 
\end{equation} 
where $\tilde{T} =  k_B T/ (N \varepsilon_L)$  is the dimensionless
temperature.  

Once equilibrium has been reached for a single simulation, we make use of 
Eq.~(\ref{eqn:fit}) to measure the quantity $\tilde{T}$.  Decomposing the wave
functions in some basis and time-averaging the populations determines $\langle
N_{k} \rangle/N$ as a function of the variable  $k$, as is plotted in 
Fig.~\ref{fig:qp_distribution}.  This completely specifies the RHS of
Eq.~(\ref{eqn:fit}) and it remains to determine the quantities on the LHS. 

In this section we consider three different methods of either predicting or
measuring the function $\tilde{\varepsilon}_{k} - \tilde{\lambda}$.  If the basis
we have used for our decomposition is a good one, and our prediction for 
$\tilde{\varepsilon}_{k} - \tilde{\lambda}$ is correct, then this curve will have
the same shape as the RHS of Eq.~(\ref{eqn:fit}).  The constant of proportionally
determined by a fitting procedure will then give the temperature $\tilde{T}$.

Before we describe our methods and results, we would like to note that the quantity
we refer to throughout the remainder of this paper as the temperature is the
variable $\tilde{T}$ as determined by the numerical fitting procedures described
above.  We have not yet established that this is the true temperature as defined by
thermal equilibrium with a heat reservoir.  However, we believe that if we were to
solve the FTGPE with $\hat{\eta}(\mathbf{x})$ acting as a heat bath, then the
temperature determined in the coherent region via this method would agree with the
bath temperature.

\subsection{Method 1 : Bogoliubov theory}

In the limit of large condensate fraction $\langle N_0 \rangle/N \sim 1$, we
expect the Bogoliubov transformation to provide a good basis.  
For the homogeneous case the dispersion relation is analytic, and is given by
\begin{equation}
\varepsilon_k - \lambda = \left[\left(\frac{\hbar^2 k^2}{2 m}\right)^2
+ (c \hbar k)^2\right]^{1/2},
\label{eqn:bog_disperse}
\end{equation}
where $c = (n_0 U_0 / m)^{1/2}$ is the speed of sound 
and $\varepsilon_k$ is the
absolute energy of a mode with wave vector $k$.  In our dimensionless
units this becomes
\begin{equation}
\tilde{\varepsilon}_{k} - \tilde{\lambda} = \left(\tilde{k}^4 + 2 
C_{\rm nl}\frac{\langle N_0 \rangle}{N} \tilde{k}^2\right)^{1/2}.
\label{eqn:bog}
\end{equation}
The condensate fraction is determined from the numerical results, and so
when the Bogoliubov dispersion relation is valid 
we can determine a temperature for the simulations
by substituting  Eq.~(\ref{eqn:bog}) in Eq.~(\ref{eqn:fit}).

\begin{figure}\centering
\includegraphics[width=8.6cm]{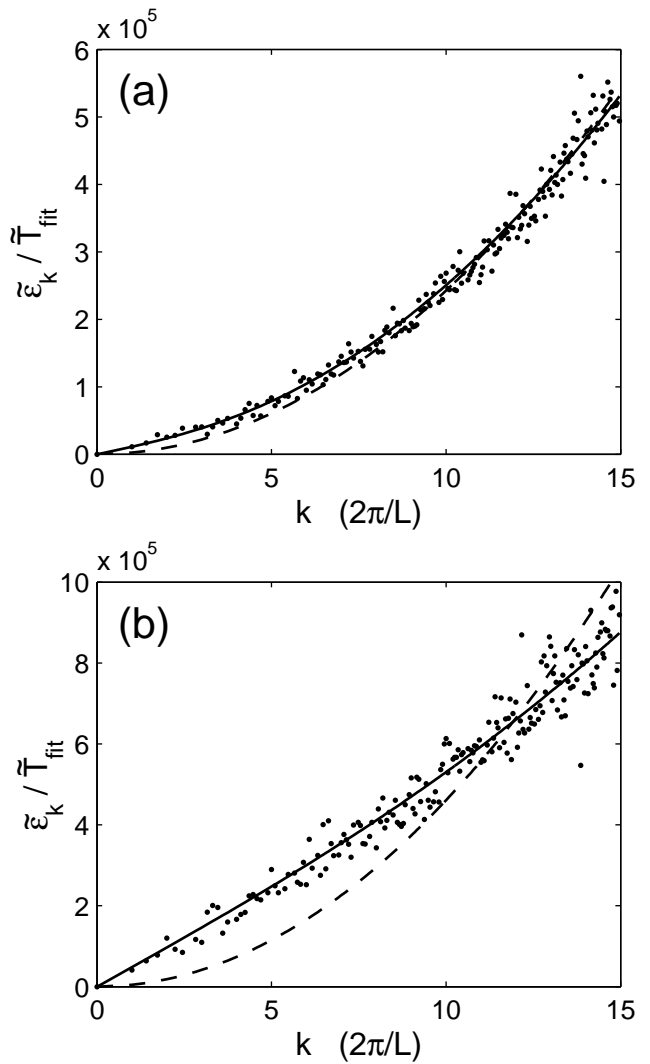}
\caption{Fits of the simulation quasiparticle 
population data to the Bogoliubov dispersion
relation for two cases.  For both graphs the solid line is the Bogoliubov
curve, while the dashed line is the ideal gas dispersion relation.  The
temperature  is determined by a least-squares fit to the plot of 
$(N/\langle N_k \rangle  - N/\langle N_0\rangle)$,
which is shown as the dots.
(a) $\Cnl = 500$, $\tilde{E} = 500$ and 
$\langle N_0\rangle/N = 0.929$, with a best fit temperature from Bogoliubov
theory of
$\tilde{T} = 0.0175$.  (b) $\Cnl = 10000$, $\tilde{E} = 5250$ and 
$\langle N_0\rangle/N = 0.957$, with a best fit temperature from Bogoliubov
theory of
$\tilde{T} = 0.018$. }
\label{fig:bog_dispersion}
\end{figure}

\subsubsection*{Results}

We have carried out this analysis for all the simulation data. For the $\Cnl =
500$ case, the measured distributions are in excellent agreement with the
Bogoliubov dispersion relation for all energies, and we have been
able to extract the corresponding temperature for each simulation.

However, this is not the case for the more strongly interacting systems.  For 
$ \Cnl = 2000$, the Bogoliubov relation is a good fit only for simulations with
$\tilde{E} \leq 2000$ ($\langle N_0 \rangle / N \geq 0.75$), or for energies
above the BEC transition point.  For the $ \Cnl = 10000$ case, good agreement is
found only for the lowest energy simulation with $\tilde{E} = 5250$ and
$\langle N_0 \rangle / N \approx 0.96$.  Sample fits of the simulation data to
the Bogoliubov dispersion relation are shown in Fig.~\ref{fig:bog_dispersion}
for cases where the agreement is good.
(An example of this procedure for where the Bogoliubov spectrum is not
appropriate is given in Fig.~\ref{fig:full_theory}).

The reason for the limited range of agreement is because the Bogoliubov
transformation diagonalises only  a quadratic approximation to the full Hamiltonian.  It neglects terms that are
cubic and quartic in non-condensate operators, assuming that they are small (these are discussed in
detail below).  This is a good approximation for the  $\Cnl = 500$
simulations---at large condensate fraction the dispersion relation is only
slightly shifted from the non-interacting relation $\tilde{\varepsilon}_k =
\tilde{k}^2$, and at smaller condensate fractions the difference is
negligible.  Hence we can fit a temperature up to and above the BEC transition.

For the $ \Cnl = 2000$ case the higher order terms  become important
above $\tilde{E} = 2000$, and for the strongest interaction strength of  $\Cnl
= 10000$, they are important for all but the lowest energy simulation we
consider.
 For the higher energy simulations the shape
of Eq.~(\ref{eqn:fit}) no longer agrees with Eq.~(\ref{eqn:bog}), and we must
use a more sophisticated theory to predict the dispersion relation.

Above the transition point, however, there is no condensate and the ideal gas
dispersion relation is a reasonable description of the system.

\subsection{Method 2 : Second order theory}

As the occupation of the quasiparticle modes becomes significant at large
interaction strengths, the cubic and quartic terms of the many-body Hamiltonian
that are neglected in the Bogoliubov transformation become important.  In
Ref.~\cite{sam} Morgan develops a consistent extension of the Bogoliubov theory
to second order that leads to a gapless excitation spectrum.  This theory
treats the cubic and quartic terms of the Hamiltonian using perturbation theory
in the Bogoliubov quasiparticle basis.  This results in energy-shifts of the
excitations away from the Bogoliubov predictions of 
Eq.~(\ref{eqn:bog_disperse}).

Expressions for the
energy-shifts of the excitations are given in Sec.~6.2 of Ref.~\cite{sam}. 
They have the form
\begin{equation}
\Delta\tilde{\varepsilon}_k = 
\Delta\tilde{E}_3(k) + \Delta\tilde{E}_4(k) + \Delta\tilde{E}_{\lambda}(k),
\end{equation}
where $\Delta\tilde{E}_3(k)$ [$\Delta\tilde{E}_4(k)$] is the shift in energy of a
quasiparticle in mode $k$ due to the cubic
[quartic] Hamiltonian, and $\Delta\tilde{E}_{\lambda}(k)$ describes the shift due to the
change in the condensate eigenvalue.  In the high-occupation limit we find
\begin{equation}
\Delta\tilde{E}_4(k) + \Delta\tilde{E}_{\lambda}(k)
= - \Cnl \tilde{\kappa} \frac{(1 + \alpha_k)^2}{ 1 - \alpha_k^2},
\label{de1}
\end{equation}
where $\tilde{\kappa}$ is the dimensionless anomalous average, defined by
\begin{equation}
\tilde{\kappa} = \sum_k \frac{(N_k + N_{-k})\alpha_k}{N(1-\alpha_k^2)}.
\label{eqn:kappa}
\end{equation}

The expression for $\Delta\tilde{E}_3(k)$ is derived from second-order
perturbation theory, and is rather complicated.  We have
\begin{equation}
\Delta\tilde{E}_3(k) =
\frac {- 2 \Cnl}{1 - \alpha_k^2}
[\Delta\tilde{E}_3^a(k) + \Delta\tilde{E}_3^b(k) + \Delta\tilde{E}^c_3(k)],
\label{de3}
\end{equation}
where
\begin{widetext}
\begin{eqnarray}
\Delta\tilde{E}_3^a(k)
&=&
\sum_j \frac{
(N_i + N_j) ( 1 - \alpha_i - \alpha_j +  
\alpha_i  \alpha_k +  \alpha_j  \alpha_k - \alpha_i  \alpha_j \alpha_k)^2}
{N(z_i + z_j - z_k)(1-\alpha_i)^2(1-\alpha_j)^2},\label{dea}
\\
\Delta\tilde{E}_3^b(k)
&=&
\sum_j \frac{
(N_{-i} + N_{-j}) ( \alpha_i + \alpha_j + \alpha_k  
 - \alpha_i  \alpha_j -  \alpha_i  \alpha_k - \alpha_j \alpha_k)^2}
{N(z_i + z_j + z_k)(1-\alpha_i)^2(1-\alpha_j)^2},
\label{deb}
\\
\Delta\tilde{E}_3^c(k)
&=&
\sum_j \frac{
(N_{i} - N_{j}) (1 -  \alpha_j -  \alpha_k  
 + \alpha_i  \alpha_j +  \alpha_i  \alpha_k - \alpha_i \alpha_j \alpha_k)^2}
{N(z_i - z_j + z_k)(1-\alpha_i)^2(1-\alpha_j)^2},
\label{dec}
\end{eqnarray}
in which ${\bf i} = {\bf k} - {\bf j}$, and
\begin{eqnarray}
z_k \; =  \; y_k (2 + y_k^2)^{1/2}
\;\equiv \; 
\tilde{\varepsilon}_k \left(\Cnl \frac{\langle N_0 \rangle}{N}\right)^{-1},
\end{eqnarray}
is another form of the
dimensionless energy of mode $k$, with $y_k = k /k_0$ as earlier.
\end{widetext}

\subsubsection{Calculation of energy shifts}

The numerical calculation of the energy shifts is not a trivial task, and  have
used two methods to determine the shifts for our simulations.  The first
procedure is to calculate the shifts directly using the population data from
the simulations. We therefore
\begin{enumerate}
\item{Calculate the quasiparticle populations $N_{\bf k}$ 
for the last 50 wave functions of our simulation based on a condensate
population $\langle N_0 \rangle $, and then
average these over time.}
\item{Calculate the energy shifts for mode ${\bf k}$ 
 using these populations as the input.}
\item{Average the shifts over angle to give a one-dimensional function of $k$.}
\end{enumerate}
This results in plots of the energy shifts that are somewhat scattered due to the
finite size of the system. The expressions for the shifts
Eqs.~(\ref{dea}--\ref{dec}) contain poles when energy matches occur, and hence the
numerical calculation is performed using an imaginary part in the denominator. 
The size of this imaginary part does not affect the shape of the curve in the
limit that it is small, but it  does affect the amount of scatter in the shifts.  
We have performed sample calculations allowing $L$ to increase while keeping other
parameters of the system constant, and this makes the curve smoother.

\begin{figure}\centering
\includegraphics[width=8.6cm]{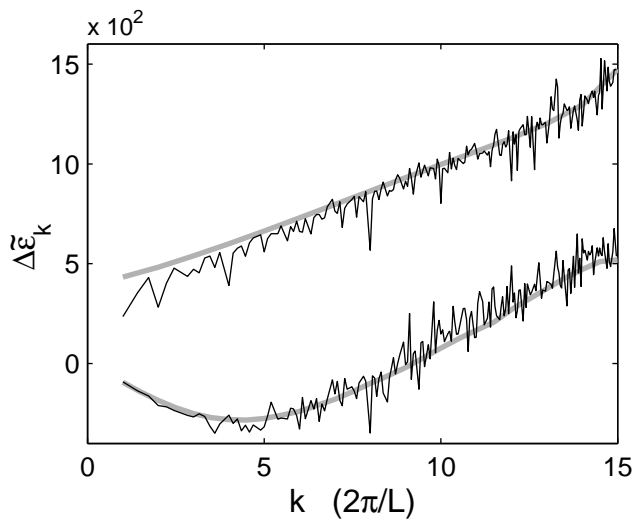}
\caption{The shifts to the Bogoliubov quasiparticle energies for two different
simulations. The solid thin curves are calculated via the first method described
in the text using population data extracted from the simulations, and are hence
somewhat noisy.  The thick grey curves use the second method, assuming
equilibrium populations given by Bogoliubov theory and calculated by numerical 
integration. The lower curves are for
the $\Cnl = 2000$, $\tilde{E} = 4000$ simulation, and appear to be
approximately gapless as  $k\rightarrow 0$. The upper curves are for the
$\Cnl = 10000$, $\tilde{E} = 6000$ simulation, and exhibit a gap as 
$k\rightarrow 0$.}
\label{fig:shift}
\end{figure}

The second procedure only makes use of the condensate fraction and the
\emph{total} number of quasiparticles, rather than the population of the
individual levels.  By assuming the Bogoliubov spectrum is a good estimate of
the energies (which must be true for the perturbation theory to be valid),  we
can  estimate the temperature $\tilde{T}_{\rm est}$ using the normalisation constraint
on the populations
\begin{equation}
\sum_k \frac{\langle N_k \rangle}{N} = \frac{\langle N_0 \rangle}{N}
+ \sum_{k>0} \frac{\tilde{T}_{\rm est}}{\tilde{\varepsilon}_k - \tilde{\lambda}},
\end{equation}
where we have used the approximation  $\tilde{\mu} = \tilde{\lambda}$ that is
valid when
there is a condensate present.  The LHS as well as the value of $\langle
N_0\rangle /N$ are determined by the simulations, and the Bogoliubov relation
Eq.~(\ref{eqn:bog}) is used for the energies.

Once the estimated temperature $T_{\rm est}$ is determined, we use the
equipartition Bogoliubov relation
for the populations in Eqs.~(\ref{dea}--\ref{dec}), and then 
approximate the sums by numerical integration to calculate the shifts to
the levels.  We find that this gives curves that agree on average with those
calculated using the first method, but are much smoother.  A comparison of the
two methods is given in Fig.~\ref{fig:shift}.

\subsubsection{Results}

 For the $\Cnl = 2000$ simulations, the quasiparticle populations extracted
from the simulations are in much better agreement with the energy spectrums
from the second order theory than with those from ordinary Bogoliubov theory.
We find that most of the measured distributions for the $\Cnl = 2000$ case
are well described by the second order theory.  Sample results are presented
in  Fig.~\ref{fig:full_theory}(a).

However,  this is not the case for the $\Cnl=10000$ simulations.  In
fact we find that the energy spectrum is shifted in the opposite direction to
that inferred from the simulations, and that there is a energy gap for
$k\rightarrow 0$.  The reasons for this are discussed below.

\subsubsection{Breakdown of perturbation theory}

\begin{figure}\centering
\includegraphics[width=8.6cm]{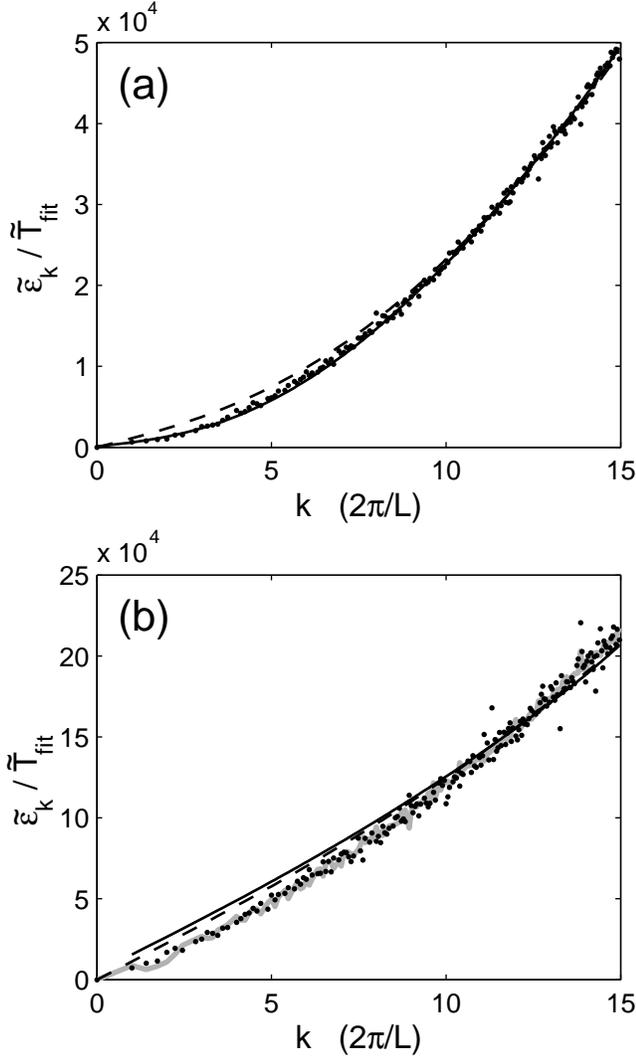}
\caption{Fits of the simulation quasiparticle  population data to dispersion
relations.  The dots are a plot of $(N/\langle N_k \rangle - N/\langle N_0
\rangle)$,  the solid curve is for dispersion relation predicted by second
order  theory, and the dashed curve is the dispersion relation predicted by
Bogoliubov theory.
(a) $\Cnl = 2000$, $\tilde{E} = 4000$,
and $\langle N_0 \rangle/N = 0.279$.  Second order theory gives a good fit to
the numerical results with a best fit temperature of $\tilde{T} = 0.201$.
(b) $\Cnl = 10000$, $\tilde{E} = 6000$,
and $\langle N_0 \rangle/N = 0.841$.  
The shape of the second order theory dispersion relation does not agree with
the population data from the simulation, and the gap is apparent as 
$k\rightarrow0$.  The grey curve plots the energies
as determined by the method described in Sec.~\ref{nonpert} with a best fit
temperature of $\tilde{T} = 0.0726$.} 
\label{fig:full_theory}
\end{figure}

The validity of the second order theory 
is constrained by the requirement \cite{sam}
\begin{equation}
\left(\frac{k_B T}{n_0 U_0}\right)(n_0 a^3)^{1/2} \ll 1,
\label{eqn:validity}
\end{equation}
where $n_0$ is the condensate density.  This corresponds 
in our dimensionless units to
\begin{equation}
\frac{\tilde{T}}{(8 \pi)^{3/2}} 
\left(\frac{C_{\rm nl}}{\langle N_0 \rangle/ N}\right)^{1/2} \ll 1.
\label{eqn:validity_dimless}
\end{equation}
For the results of Fig.~\ref{fig:full_theory} with $\Cnl = 2000$, 
$\tilde{E} = 4000$ this
parameter is 0.14 and so we are beginning to probe the boundary of validity of
the theory.   At higher $\tilde{E}$ the shifts become of the order of the
unperturbed energies, and hence the results are unreliable. In this region even higher
order terms are important, and the second order theory can no longer be
expected to give good results.  From our calculations it seems that this
parameter should be $\leq 0.2$ for the theory to be valid.

We would like to emphasize, however, that the GPE suffers no such
limitations.  It is non-perturbative and thus we expect  that it will be valid all
the way through the transition region as long as the high occupation number
condition is satisfied.

\subsubsection{Gaplessness in a finite system}
In the course of this work it has become apparent that while the second order
theory is gapless for infinite systems, this is not the case for systems such
as ours
with a finite momentum cutoff.  The individual terms in the perturbation
expansion given by Eqs.~(\ref{de1}) and (\ref{de3}) contain contributions that
are proportional to $1/k$
(infrared divergent) and a constant (gap) in the low $k$ limit.  For a
homogeneous system
these terms cancel exactly when the upper limit of the integrals is infinite,
and this leaves a gapless spectrum \cite{sam,fedichev,giorgini}.  However, in a
system with a momentum cutoff these terms do not exactly cancel, with the
result that there is a gap in the predicted excitation spectrum as $k
\rightarrow 0$. 

Briefly, this gap arises because the energy shifts
$\Delta\tilde{E}_4(k)+\Delta\tilde{E}_{\lambda}(k)$ of Eq.~(\ref{de1}) only
involve the quantity $\tilde{\kappa}$. This is obtained from
Eq.~(\ref{eqn:kappa}) via a sum over all states below the cutoff where the
summand depends only on a single wavevector. In contrast the shift
$\Delta\tilde{E}_3(k)$ of Eq.~(\ref{de3}) involves a sum over states where the
summand depends on two wavevectors ${\bf i}$ and ${\bf j}$ (related by momentum
conservation) \emph{both} of which must be below the cutoff. This difference in
the restrictions on the summations leads to a lack of complete cancellation in
the corresponding shifts at low energy and the appearance of a gap in the
excitation spectrum.

For the homogeneous gas, we can calculate the size of the gap predicted by the
second order theory analytically. Replacing the summations by integrations, we
find that the leading order contribution to the energy shift in the limit $k
\rightarrow 0$ is
\begin{equation}
\Delta \epsilon_k = \left(\frac{k_B T}{n_0 U_0}\right)(n_0 a^3)^{1/2}
\left(\frac{8}{\pi}\right)^{1/2} \frac{\epsilon_k}{y_k(2+y^2_c)} +
\mbox{O}(y_k).
\label{dek_gap}
\end{equation}
where $y = k/k_0$ as before and $y_c = k_c/k_0$ where $k_c$ is the momentum
cutoff.
In the limit $k \rightarrow 0$ we have $\epsilon_k \propto y_k$ and so $\Delta
\epsilon_k$ tends to a constant (the gap). The size of the gap tends to zero as
the momentum cutoff $y_c$ tends to infinity but otherwise it is finite. We
stress that Eq.~(\ref{dek_gap}) is only the low $k$ limit of the exact result.
For our simulations there is a minimum wavevector in the problem so it is
possible for the terms of order $y_k$ to be larger than the gap contribution
given above. This is the case for the simulations with $\Cnl \leq 2000$ where
$y_c$ is reasonably large and the gap is therefore small.

The result of Eq.~(\ref{dek_gap}) contains the small parameter that controls
the validity of the second order theory in the usual case where there is no
momentum cutoff [c.f. Eq.~(\ref{eqn:validity})]. However, the result also
depends explicitly on the cutoff $k_c$ so in this case there is a second
parameter in the theory. For perturbation theory to be valid we require that
the predicted energy shifts are small compared to the unperturbed energies,
i.e. that $\Delta \epsilon_k/\epsilon_k \ll 1$. We therefore obtain a second
criterion for the validity of the second order theory which is
\begin{equation}
\left(\frac{k_B T}{n_0 U_0}\right)(n_0 a^3)^{1/2}
\left(\frac{8}{\pi}\right)^{1/2} \frac{1}{y_k(2+y^2_c)} \ll 1.
\end{equation}
This result should hold for all momenta in the simulations, and in particular
for the smallest value of $y_k$. For the $\Cnl = 2000$, $\tilde{E} = 4000$
simulations the left hand side is $0.04$ for $\tilde{k} = 2\pi$. In this case
the gap is negligible and the dominant contribution to the energy shifts comes
from the terms of order $y_k$ in Eq.~(\ref{dek_gap}). The small parameter of
the theory is therefore given by Eq.~(\ref{eqn:validity}). However, for the
$\Cnl = 10000$, $\tilde{E} = 5250$ simulations the left hand side is of order
0.12. In this case the gap is not negligible and we cannot use second order
theory to define a temperature.

This result is somewhat surprising since, even for a condensate fraction of
80\% the small parameter of Eq.~(\ref{eqn:validity}) is of order 0.07, and it
does seem
reasonable to expect that perturbation theory should be applicable. This does
not appear to be the case, however, and we have so far been unable to determine
the root cause of this problem. It is worth noting that the numerical
simulations themselves have no difficulties in this regime and do not predict a
gap at low momentum. This is because the GPE is non-perturbative and indeed
this is one of the main reasons for using it to study the properties of Bose
condensed systems at finite temperature.

The disagreement between the second order theory and the numerical simulations
is illustrated in Fig.~\ref{fig:full_theory}(b), where it can be seen that even
despite the gap, the shifts the theory predicts are in the wrong direction in
comparision with the simulations.

\subsection{Method 3 : Non-perturbative determination of the temperature}
\label{nonpert}

The failure of second order theory for the $\Cnl = 10000$ simulations caused us
to investigate other possible methods of determining the temperature once the
system was in equilibrium.  This has lead to what seems to be a method of
determining the temperature that does not rely on perturbation theory, and we
describe it here.

We found  earlier that the Bogoliubov spectrum gave a good prediction of the
populations of the quasiparticle levels for the lowest energy simulation in
the  $\Cnl = 10000$ series with $\tilde{E} = 5250$.  Therefore it seems
reasonable that the Bogoliubov \emph{basis} should remain a good one for
perturbation theory for the next simulation with $\tilde{E} = 5500$, even
though the second order theory cannot be used to calculate the energy shifts.

Therefore we attempted another method to determine the absolute energy of each
quasiparticle level.  If we are using a good basis, then on a short time scale
the quasiparticles should be independent, with amplitudes evolving according to
\begin{equation}
b_{\bf k}(t) = b_{\bf k}(t_0) \exp(-i \varepsilon_{\bf k} (t - t_0) / \hbar).
\end{equation}
Thus by measuring the gradient of the phase of each quasiparticle we can determine its
energy.  

To determine the energy spectrum for a single simulation, our numerical
procedure was as follows.

\begin{enumerate}
\item
Take the last 50 wave functions saved for a simulation once it has reached
equilibrium, and evolve each of these individually 
for a very short period.  One hundred wave functions are saved for each of
the 50 simulations.
\item
 Transform the wave functions into the
quasiparticle basis, and measure the energy of each quasiparticle, determined
 by a linear fit to the phase of each amplitude for all 50 simulations.
\item 
 Average
over all 50 energy spectrums to give a single three dimensional spectrum.
\item  Finally, average over angle
to give a one dimensional energy spectrum.
\end{enumerate}

This gives us a dispersion relation $\tilde{\varepsilon}_k-
\tilde{\lambda}$ which can then be compared to a plot of  
$(N/\langle N_k \rangle  - N/\langle N_0\rangle)$.  If the shapes of the curves
agree, then a temperature can be determined via Eq.~(\ref{eqn:fit}) as in the
earlier sections.

We first tested this procedure on the  $\Cnl = 2000$ simulation series, and
found that this method was in good agreement with the second order theory
calculations, the two approaches assigning the same temperature to the various
simulations.

We then moved onto the  $\Cnl = 10000$ simulations.  We found the surprising
result that not only did the shape of the plots of the curves for
$\varepsilon_k / k_B T_{\rm fit}$ and $(1/N_k - 1/N_0)$ agree for the lower 
energy simulations where the parameter of Eq.~(\ref{eqn:validity}) was small,
it also agreed when it was of the order of, and greater than one.  This was
unexpected, as it would seem likely that near the phase transition when
interactions are strong that the Bogoliubov quasiparticle basis would no longer
be sufficiently good for this method to be accurate.  An example of the energy
spectrum and its fit to the population data is shown in
Fig.~\ref{fig:full_theory}(b).

As a further test we carried out the same procedure  described above, but using
the plane wave basis rather than the Bogoliubov quasiparticle basis. 
Intuitively it would seem that this would no longer work---but we found that
not only did it give the same temperatures as the quasiparticle basis for the 
$\Cnl = 10000$ simulations, it also agreed with the temperatures determined
using second order theory for the $\Cnl = 2000$ simulations.

\section{Temperature dependence}\label{frac_vortex}

Using the three methods described in the previous section, we have been able to
measure an equilibrium temperature for \emph{all} simulations in this paper.  We
are confident of the results determined from both Bogoliubov and second order
theory; however unfortunately we do not have any results to compare with for the
strongly interacting regime where the non-perturbative method was used.  We can
only conclude that the temperatures extracted using this method agree numerically
with the other two methods in the weakly interacting regime, and that the
values obtained seem reasonable and basis independent elsewhere.   We intend to
test this method further in the future using a numerical ``ideal gas thermometer''.

In this section we move on to consider how other system properties such as
condensate fraction, specific heat, and vorticity vary with the temperature
$\tilde{T}$.

\subsection{Condensate fraction}

\begin{figure}\centering
\includegraphics[width=8.6cm]{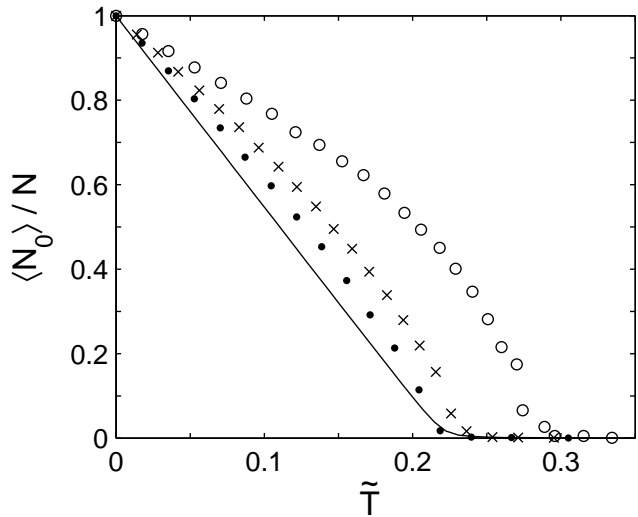}
\caption{Condensate fraction versus temperature for the PGPE system with 
$k < 15 \times 2 \pi/L$ for
four different interaction strengths.  
The open circles are for $C_{\rm nl} = 10000$,
crosses for $C_{\rm nl} = 2000$, solid dots for $C_{\rm nl} = 500$,
and the solid line is for the ideal gas.
The shift in the transition temperature is positive with increasing interaction
strength $C_{\rm nl}$.
} 
\label{fig:n0_v_T}
\end{figure}

It is usual when considering how the condensate fraction varies with the other
properties of the system to plot it against temperature, rather than against
energy as we have done in Fig.~\ref{fig:n0ve}.  We are now in a position to
present this data, and it is displayed in Fig.~\ref{fig:n0_v_T}.
We can see that a major effect of  increasing the nonlinearity is to increase
the condensate fraction at any given temperature.  This can be understood in
the Bogoliubov regime by considering the shape of the dispersion relation.

The Bogoliubov dispersion relation Eq.~(\ref{eqn:bog})  shows that for a given
condensate fraction, a larger value of $\Cnl$ will result in an increase in the
energy of any mode $k$  relative to the condensate.  This leads directly to the
observation that for a fixed condensate fraction, an increase in the
nonlinearity must lead to an increase in the temperature. However, as $\langle
N_0 \rangle /N \rightarrow 0$ in the transition region, the energy-momentum
relationship tends towards the ideal gas dispersion relation, and therefore the
transition temperature will not be greatly shifted over a wide range of
nonlinearities.

There has been some discussion recently in the literature about the shift in
the transition temperature for the homogeneous interacting Bose gas, with some
authors even disagreeing in the direction of the shift (e.g. see
Ref.~\cite{arnold} and references within).  For the PGPE system described in
this paper by Eq.~(\ref{eqn:pgpe}), we can see from Fig.~\ref{fig:n0_v_T} that
the shift is positive, although as yet we have made no effort to quantify
this.  This would require many more simulations to be run, especially in the
transition region, and for the temperatures to be determined more accurately. 

It seems plausible that  future simulations of the full FTGPE (\ref{eqn:ftgpe}) or
approximations to it could be used to quantitatively measure the shift in the
critical temperature for the homogeneous Bose gas when the lowest energy modes are
sufficiently classical.  However, the terms coupling the FTGPE to the effective
bath $\hat{\eta}(\mathbf{x})$ may be difficult to implement computationally, and at
the present time we are unsure how to proceed in this direction.

\subsection{Specific heat}

\begin{figure}\centering
\includegraphics{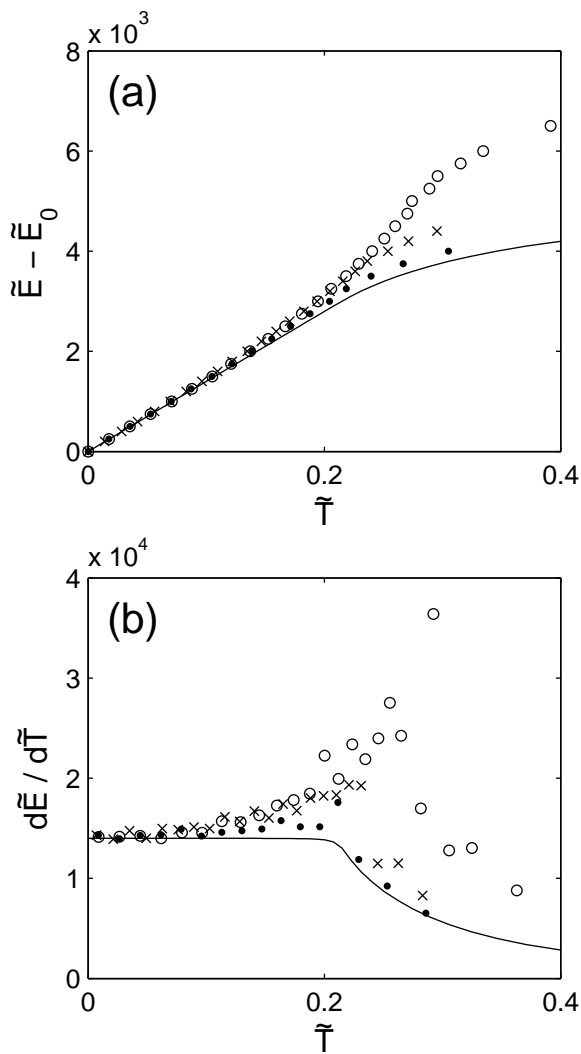}
\caption{
Graphs relating to the specific heat of the Bose gas in the PGPE model.
  The open circles are for
$C_{\rm nl} = 10000$, crosses for $C_{\rm nl} = 2000$, solid dots for $C_{\rm
nl} = 500$, and the solid line is for the ideal gas.
(a) Plot of the energy versus temperature for all four interaction
strengths considered.  (b) Plot of the specific heat for all four interaction
strengths.} 
\label{fig:specific}
\end{figure}

In Fig.~\ref{fig:specific}(a) we plot the energy of the simulations due to 
 excited states $(\tilde{E} - \tilde{E_0})$  versus temperature,
where  $\tilde{E}_0 = \Cnl/2$ is the energy of the system at $\tilde{T} = 0$.
We can see that at low temperatures for all interaction strengths this is a
straight line, with a slope of about 13996---the number of excited modes in the
system. This is as expected---the average 
energy contained in a given mode $k$ is 
\begin{eqnarray}
\langle\tilde{E_k} \rangle \equiv \frac{\langle N_k\rangle}{N}
\tilde{\varepsilon}_k=\frac{\tilde{T}}{\tilde{\varepsilon}_k - \tilde{\mu}}
\tilde{\varepsilon}_k\approx  \tilde{T},  \end{eqnarray}  when there is a
condensate present and $\tilde{\mu} \rightarrow 0^-$. At higher temperatures,
however, the energy rises above the equipartition  prediction for non-zero
interaction strength $\Cnl$, and this is an indication that there are no longer
independent modes in the system.

The derivative of this curve with respect to temperature gives the specific
heat, and this quantity is plotted in Fig.~\ref{fig:specific}(b).  The messy
nature of this plot is due to small uncertainties in the measured temperature
which are amplified when the temperature difference between successive
simulations is calculated.  However, the plot does display an interesting
feature.  For non-zero interaction strength, the specific heat appears to reach
a peak at the transition temperature, and the height of this peak increases
with the value of $\Cnl$---somewhat reminiscent of the lambda transition in
superfluid helium. Once again, further simulations and more accurate
determination of the temperature are required for quantitative investigation of
this effect.  This will be the subject of future work.

\subsection{The role of vortices}

A further quantity of interest is the vorticity of the system in equilibrium. 
It has been argued that vortices may be important in the superfluid transition
of $^4$He, reducing the superfluid density near the transition point
\cite{Williams}.  With this in mind, we have studied the presence of vortex
lines in our simulations.  Recently Berloff and Svistunov \cite{berloff} have
considered the evolution of topological defects in the evolution of a Bose gas
from a strongly non-equilibrium state.

A vortex  is a topological excitation, characterised in a wave function by
\begin{equation}
\oint_C \nabla \mbox{Arg}[\psi(\mathbf{x})] \cdot d{\bf l} = 2 \pi n,
\end{equation}
where $C$ is a closed contour, and $n$ is a non-zero integer, the sign of which
indicates the circulation of the vortex. 
The continuous variation of the phase from zero to $2 n \pi$
around such a contour implies that there
must be a discontinuity in the phase within the loop.  The only way
that this can be physical is for the wave function to have zero amplitude at
the spatial position of the phase singularity.

In a two-dimensional wave function the centre of vortices are zero-dimensional
points, and they can be easily counted to give a measure of the vorticity of the
system. However, in three dimensions vortices form lines and rings, and the
equivalent quantity of the 2D measure of vorticity  would be to calculate the
length of all vortex structures in the wave function. This would be a somewhat
complicated procedure numerically, and so we have devised a different
technique.

We increase the spatial resolution of our wave functions to be $128\times 128
\times 128$ points, so that the grid spacing is smaller than the vortex healing
length $\xi$, defined by
\begin{equation}
\frac{\hbar^2}{2 m \xi^2} = n_0 U_0.
\end{equation}
We do this by extending the wave function in {\em k}-space, and then
Fourier transforming to real space. This does not require any extra
information, as for $k > 15 \times 2\pi/L$ we have $c_{\bf k} = 0$.  We
then count the number of vortex lines passing through every $xy$ plane, and
take the average over all planes.  It seems that this is a reasonable
measure of the vorticity  of the wave function, and it should be similar to the
measurement of the length of the vortex structures discussed above.

We have analysed the data from the simulations using this procedure.  We find
that when the energy of the simulation is sufficiently high that there are
vortices present, the time evolution of the vorticity is a good indicator for
when the system reaches equilibrium.  As is the case for the condensate
population, the vorticity tends to an equilibrium value which fluctuates by a
small  amount (much smaller than the fluctuations in the condensate
population).

\begin{figure}\centering
\includegraphics{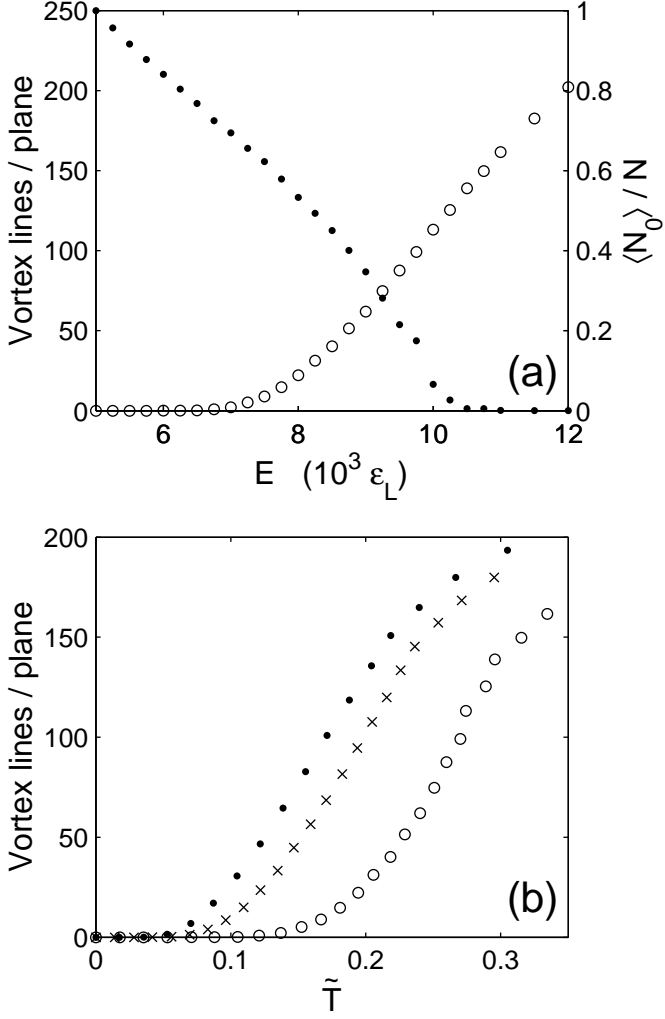}
\caption{The presence of vortices in the simulations.  (a) A plot of 
vorticity for the $\Cnl=10000$ simulation series.  The number of vortex lines
per plane are indicated by open circles with the scale on the  
left vertical axis, and the condensate fraction by dots with the scale on the
right vertical axis. (b) The number of vortex lines per plane plotted against temperature
for all three simulation series.  Open circles are $\Cnl = 10000$, crosses are 
$\Cnl = 2000$, and dots are $\Cnl = 500$. }
\label{fig:vortices}
\end{figure}

\begin{figure}\centering
\includegraphics[width=8.6cm]{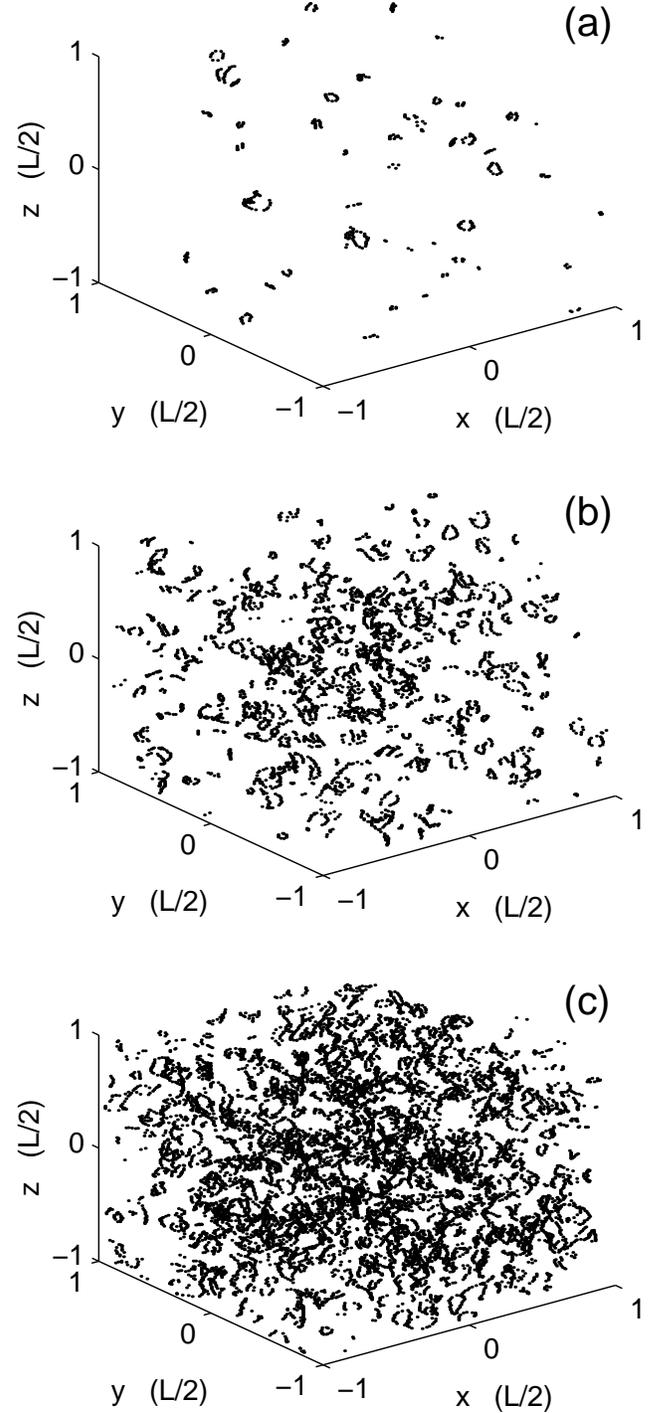}
\caption{A visualisation of the ``vortex tangle'' in equilibrium for the case of
$\Cnl = 10000$.  (a) $\tilde{E} = 7000$, (b)  $\tilde{E} = 8000$, 
(c)  $\tilde{E} = 9000$.   Each point corresponds to where a vortex line was
detected in the horizontal plane.  Several vortex rings are visible in 
the figures.}
\label{fig:vortex_tangle}
\end{figure}

A plot of the vorticity against system energy is shown in 
Fig.~\ref{fig:vortices}(a) for the $\Cnl = 10000$ simulation (the curves
are qualitatively similar for the other nonlinearities).  We see
that there is a minimum energy required for vortices to be present in the
system at equilibrium.  Also, as we reach this energy the plot
of condensate occupation versus energy appears to dip.  This same behaviour is
observed for the $\Cnl = 500$ and $2000$ cases, but it occurs at a higher
condensate fraction, and  is
not as pronounced.  There is no corresponding departure from linearity in the
ideal gas case, as was seen in Fig.~\ref{fig:n0ve}(b).

A plot of the number of vortex lines versus temperature for all the
simulations is shown in Fig.~\ref{fig:vortices}(b), and this displays a
large increase in the vorticity near the transition temperature for $\Cnl =
10000$. Even the $\Cnl=2000$ case appears to show a small jump in this
region.  A more in-depth analysis of this behaviour will be carried out in a
subsequent extension of this work.

Finally, a three-dimensional visualisation of the network of vortex lines is
shown in Fig.~\ref{fig:vortex_tangle} for three simulation energies for the 
$\Cnl = 10000$ simulations.  Each point corresponds to where a vortex line was
detected in the horizontal planes, and for the lowest two energies several
vortex rings are clearly visible.

\section{Conclusions}\label{conclusions}

We have presented what we believe is compelling evidence that the projected
\GPE is a good approximation to the dynamics of the classical modes  of a Bose
gas.  We have described how to carry out the projection technique in the
homogeneous case with periodic boundary conditions, and have shown that
starting with a randomised wave function with a given energy, the projected GPE
evolves towards an equilibrium state.  We have analysed the numerical data in
terms of quadratic Bogoliubov theory, and also the gapless, finite temperature
theory of Ref.~\cite{sam} in the classical limit.  We have found that both the
occupation and energies of the quasiparticles agree quantitatively with the
predictions when these theories are valid.

Outside the range of perturbation theory we have proposed another technique
that has allowed us to determine a temperature for the PGPE simulations in
equilibrium.  This method agrees with the perturbative methods when they are
valid.  Using this definition, we have found that increasing the nonlinearity
$\Cnl$ leads to both an increase in the transition temperature, and in the
specific heat of the system at the critical point. We have also presented
evidence that suggests vortices may play some role in the transition.  The
projected GPE is a simple equation but it appears to describe very rich
physics, only some of which we have considered here. 

\begin{acknowledgments}
The authors would like to thank R.~J.~Ballagh and C.~W.~Gardiner for useful discussions.
This work was financially supported by St John's College and Trinity College,
Oxford, and the UK-EPSRC.
\end{acknowledgments}


\begin{thebibliography}{99}

\bibitem{formalism}
M.~J.~Davis, R.~J.~Ballagh, and K.~Burnett, 
J. Phys. B \textbf{34}, 4487 (2001).


\bibitem{JILA} M.~Anderson,
J.~R.~Ensher, M.~R.~Matthews, C.~E.~Wieman and E.~A.~Cornell, 
Science {\bf 269}, 198 (1995).

\bibitem{MIT} K.~B.~Davis,
M-O.~Mewes, M.~R.~Andrews, N.~J.~van~Druten, D.~S.~Durfee, D.~M.~Kurn, and W.~Ketterle, 
Phys.  Rev.  Lett.  {\bf 75}, 3969 (1995).

\bibitem{RICE} C.~C.~Bradley,
 C.~A.~Sackett, J.~J.~Tollet, and R.~G.~Hulet, 
Phys. Rev. Lett. {\bf 75}, 1687 (1995).


\bibitem{sam} S.~A.~Morgan, J.~Phys.~B {\bf 33}, 3847 (2000).

\bibitem{fedichev}
P.~O.~Fedichev and G.~V.~Shlyapnikov, Phys. Rev. A {\bf 58}, 3146 (1998).

\bibitem{giorgini}
S. Giorgini, Phys. Rev. A {\bf 61}, 063615 (2000).


\bibitem{ftgpe} M.~J.~Davis, S.~A.~Morgan, and K.~Burnett,
Phys. Rev. Lett. {\bf 87} 160402 (2001).


\bibitem{dalfovo}
F.~Dalfovo, S.~Giorgini, L.~P.~Pitaevskii and S.~Stringari,
Rev. Mod. Phys. {\bf 71}, 463 (1999).


\bibitem{boris} B.~V.~Svistunov,
J. Moscow Phys. Soc. {\bf 1}, 373 (1991).


\bibitem{kagan1} Yu.~Kagan, B.~V.~Svistunov, and G.~V.~Shlyapnikov, Zh. Eksp.
Teor. Fiz. {\bf 101}, 528 (1992) [Sov. Phys. JETP {\bf 75}, 387 (1992)].


\bibitem{kagan2}  Yu.~Kagan and B.~V.~Svistunov, Zh. Eksp.
Teor. Fiz. {\bf 105}, 353 (1994) [Sov. Phys. JETP {\bf 78}, 187 (1994)].

\bibitem{kagan3}  
Yu.~Kagan and B.~V.~Svistunov, Phys. Rev. Lett. {\bf 79}, 3331 (1997).

\bibitem{damle} K.~Damle, S.~N.~Majumdar, and S.~Sachdev, Phys. Rev. A 
{\bf 54}, 5037 (1996).


\bibitem{Marshall} R.~J.~Marshall, G.~H.~C.~New, K.~Burnett, and S.~Choi, 
Phys. Rev. A {\bf 59}, 2085 (1999).

\bibitem{sinatra1}
A. Sinatra, Y. Castin and C. Lobo,
Jour. of Mod. Opt. \textbf{47}, 2629 (2000).

\bibitem{sinatra2} A.~Sinatra, C.~Lobo, and Y.~Castin,
Phys. Rev. Lett . \textbf{87}, 210404 (2001).

\bibitem{sinatra3} A.~Sinatra, C.~Lobo, and Y.~Castin,
cond-mat/0203259 (2002).

\bibitem{goral1} K.~G\`{o}ral, M.~Gajda, and K.~Rz\c{a}\.{z}ewski,
Opt. Express {\bf 8}, 92 (2001).

\bibitem{goral2} K.~G\`{o}ral, M.~Gajda, and K.~Rz\c{a}\.{z}ewski,
cond-mat/0203259 (2002).

\bibitem{bijlsma} 
H.~T.~C.~Stoof and M.~J.~Bijlsma,
J. Low. Temp. Phys \textbf{124} 431 (2001).

\bibitem{Turok} G.~D.~Moore and N.~Turok, Phys. Rev. D \textbf{55}, 6538 (1997).

\bibitem{reichl} L.~E. Reichl, \emph{A Modern Course in Statistical Physics},
(University of Texas Press, Austin, 1980).

\bibitem{arnold}
P. Arnold, G. Moore, and B Tom\'{a}\v{s}ik,
Phys. Rev. A \textbf{65}, 013606 (2002)

\bibitem{Williams} G.~A.~Williams, J.~Low.~Temp.~Phys.\textbf{89}, 91 (1992).

\bibitem{berloff} N.~G.~Berloff and B. V. Svistunov, 
Phys. Rev. A \textbf{66}, 013603 (2002).

\end{thebibliography}
\end{document}